# Ferrimagnetic organelles in multicellular organisms

**Gorobets S. V.**[1], **Gorobets O. Yu.**[1,2], **Gorobets Yu. I.**[2,1], **Bulaievska M. O.**[1]

[1]National Technical University of Ukraine "Igor Sikorsky Kyiv Polytechnic Institute", 03056, Kyiv, Ukraine
[2]Institute of Magnetism NAS of Ukraine and MESYS of Ukraine, 03142, Kyiv, Ukraine

**Abstract**

**Introduction**
Biogenic ferrimagnetism has been intensively studied for more than four decades in connection with the discovery of biogenic magnetic nanoparticles (BMN) in a wide variety of organisms[1-40]. In the narrow sense, BMNs are nanoparticles of strong natural magnets (ferrites – magnetite, maghemite, greigit) in organisms that create stray magnetic fields in their vicinity[41,42]. In a broad sense, BMNs are nanoparticles of iron-containing materials that have sufficient magnetic susceptibility for movement in the magnetic field of laboratory magnets[10]. These nanoparticles arecalled biogenic because their biosynthesis is programmed at the genetic level[41-43]. The main impetus for the study of biogenic ferrimagnetism was the ability of magnetotactic bacteria to move along the lines of force of the Earth's magnetic field (i.e. magnetotaxis)[1,2]. Subsequently, the idea of magnetotaxis was developed into the idea of magnetoreception of animals (i.e. their ability to orient in the geomagnetic field) due to the presence of BMN[9,33].

But when examining organs and tissues for the presence of BMN, these magnetic nanoparticles were found not only in those organs and tissues that may be responsible for the orientation of animals in the Earth's external magnetic field (brain, beak of migratory birds, lateral line and ethmoid bone of migratory fish), but also in a number of other organs, both migratory and non-migratory organisms[8,14,19,22,24,25,28-30,33,37-39,44-50] (Fig. 1, Table 1).

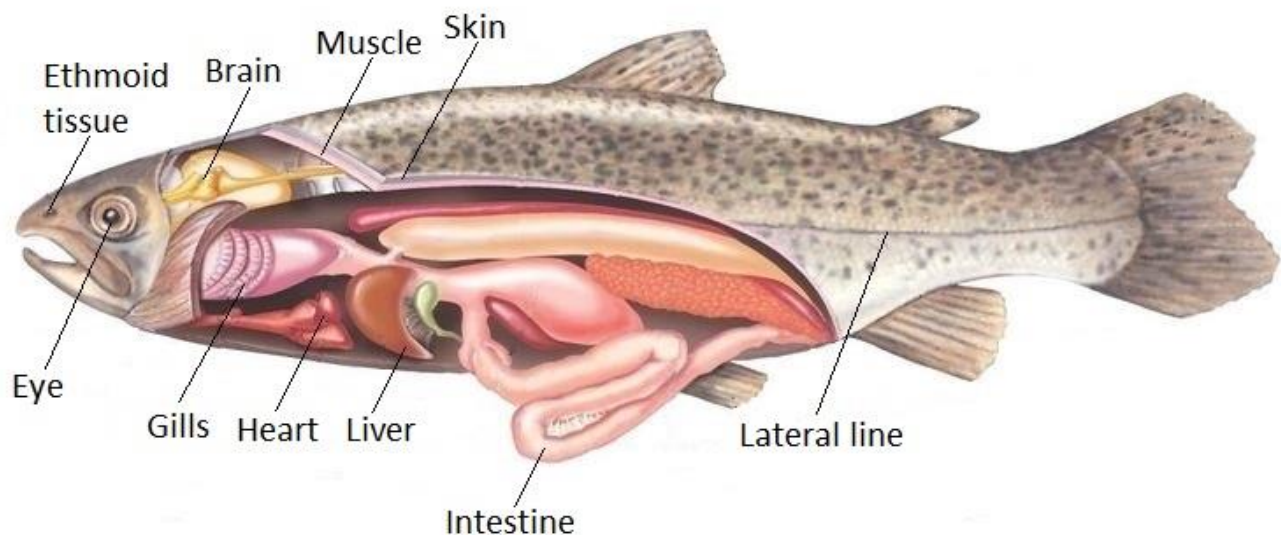

Fig. 1. The organs of migratory fish, in which the presence of BMNs is experimentally shown: ethmoid bone[22,45,46,51], brain[19,22,45], muscle[22,45,46], skin[22,45,46,51], eye[22,45,46,51], gills[45], heart[45], liver[45], intestine[45], lateral line[22,45]. The organs of non-migratory fish, in which the presence of BMNs is experimentally shown: ethmoid bone, brain, lateral line[24,25].

All these numerous experimental data on the detection of BMNs in analogous and homologous organs of organisms of different phylogenetic groups confirm the existence of the theoretically predicted single genetically programmed mechanism of biosynthesis of BMNs in all organisms[42,52,53]. At the same time, the theoretically predicted presence of BMNs in a number of human organs and tissues by the methods of comparative genomics[50]is confirmed by experimental studies on the presence of BMNs in human organs[34,37,38,54-56], as well as analogous and homologous organs and tissues of animals[8,14,19,22,25,33,44-47,49,50] (Table 1).

Table 1. Human organs, as well as analogous and homologous organs of animals, in which BMNs are found.

| Human organs with theoretically predicted BMNs presence (+) | Experimentally confirmed BMNs presence in human organs | Experimentally confirmed BMNs presence in relevant or analogous fish organs |
|---|---|---|
| Brain[50] (+) | Brain[34,54,55] | Brain of the *Oncorhynchus nerka*[19], brain of the whale *Megaptera novaeangliae*[44], head *Pachycondyla marginata*[14], brain of the silver carp (Fig. 9) |
| Heart[50] (+) | Heart[37] | Heart of the *Sus domestica*[49], fish heart[45] |
| Liver[50] (+) | Liver[37] | Liver of the *Sus domestica*[49], fish liver[45], liver of the house mouse *Mus musculus* (Fig. 2, 3) |
| Spleen[50] (+) | Spleen[37] | Spleen of the *Sus domestica*[50] |
| Ethmoid bone[50] (+) | Ethmoid bone[56] | Ethmoid bone of the *Salmo salar*, ethmoid bone of the *Cyprinus carpio*, ethmoid bone of the *Esox lucius*[25], ethmoid bone of the *Oncorhynchus nerka*[8], ethmoid bone of the *Delphinus delphis*[33], ethmoid bone of the *Thunnus albacares*[46], antennas of the *Pachycondyla marginata*[14] |
| Adrenal glands[50](+) | Adrenal glands[38] | |
| Lung[50](+) | | Lung of the *Sus domestica*[49], fish gills[45] |
| Kidney[50] (+) | | Kidney of the *Sus domestica*[50] |
| Intestine[50] (+) | | Intestine of the fish[45], abdominal cavity of the ants[14], intestine of the house mouse *Mus musculus* (Fig. 5) |
| Muscle tissue[50](+) | | Muscle of the fish[22,46] |
| Skin[50] (+) | | Skin of the *Salmo salar*[22], skin of the *Oncorhynchus nerka*[47], skin of the *Thunnus albacares*[46] |
| Eyes | | Eyes of the fish[22,45] |
| Joints | | Joints of the *Pachycondyla marginata*[14] |

In this work, the localization of BMNs was investigated using atomic force (AFM) and magnetic force microscopy (MFM) in a number of organs of various organisms (animals, plants, fungi) in order to identify which systems of multicellular organisms include BMNs.

**Experiment and discussion**
In this paper, the following samples were examined using AFM and MFM, namely, animal samples: liver, intestine, pancreas of mouse, lungs, kidneys, spleen of pig, brain of carp; plant samples: leaves and root of tobacco, stem and tuber of potato; and fungi samples: fruit bodies of the common and shiitake mushrooms. The cells and other structural components (shown in the images by arrows) in the samples of the listed organs were identified and the localization of BMNs in the indicated organs was shown.

The results of studies of samples of the liver of a mouse with the use of AFM and MFM showed that BMNs in the liver of a mouse are located in the wall of sinusoids (capillaries of the liver) (Fig. 2-3 and Table 2). Sinusoids[57], sinusoidal endothelial cells of the liver[58], nuclei of sinusoidal endothelial cells of the liver[57,59], hepatocytes, nuclei of hepatocytes[57], fenestrae[58-61] presented in Fig. 2-3 in this paper have typical morphology and dimensions, characterized in the works[57-61].

Table 2. Dimensional characteristics of the structural components of the mouse liver

| Structural components | Calculated dimensions (see Fig. 2-3), μm | Dimensions in literature, μm | References |
|---|---|---|---|
| Capillary density | 400-600/μm² | 400-500/μm² | 62 |
| Sinusoids | 5-20 | 5-15 | 63-65 |
| Sinusoidal endothelial cells | 15-25 | 20-25 | 58 |
| Hepatocytes | 15-25 | 20-30 | 58 |
| Hepatocyte nuclei | 5-8 | 6-8 | 58 |
| Fenestrae | 200-500 | 50-500 | 58-61 |

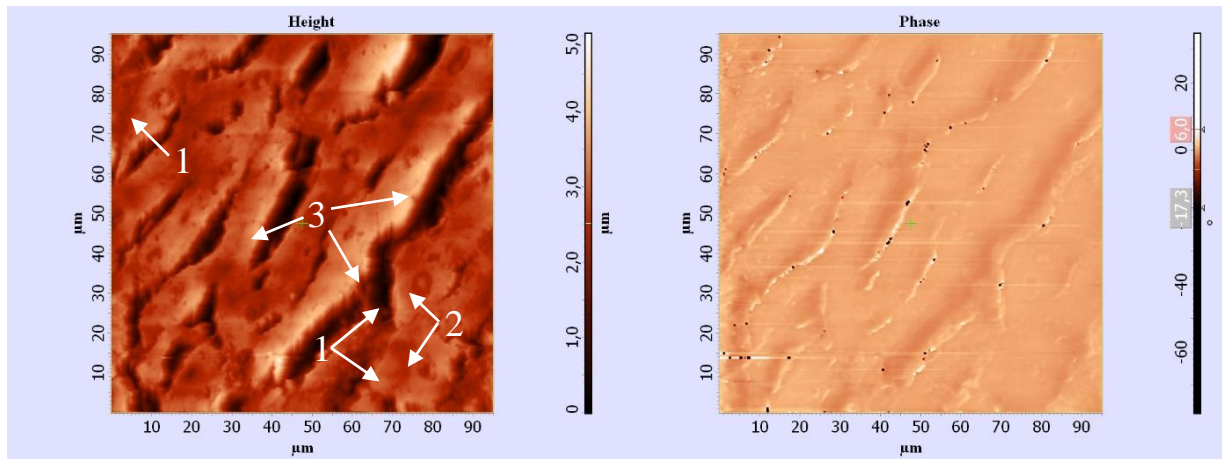

Fig. 2. AFM (left) and MFM (right) images of a sample of the liver of the house mouse *Mus musculus*: 1 hepatocyte, 2 nuclei of hepatocytes and 3 sinusoids.

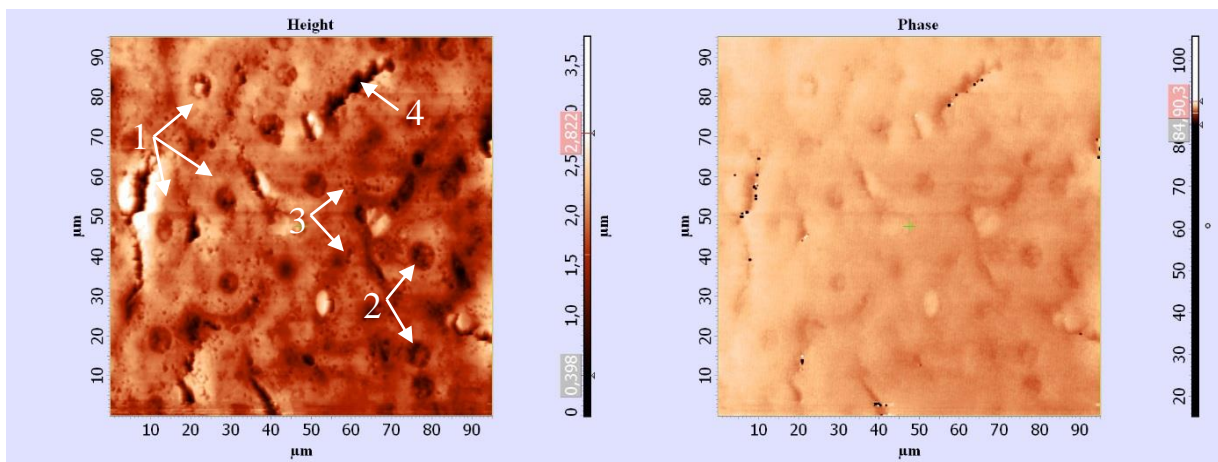

Fig. 3. AFM (left) and MFM (right) images of a sample of the liver of the house mouse *Mus musculus*: 1 sinusoidal endothelial cells of the liver, 2 nuclei of sinusoidal endothelial cells of the liver, 3 fenestrae and 4 sinusoids.

Similar results were obtained on the localization of BMNs in the samples of the pancreas and intestines of a mouse. So, BMNs in the pancreas and the intestines of the mouse are located in the wall of the capillaries as can be seen from Figures 4-5 and data on the size and distance between the capillaries in Tables 3-4. The capillaries of the pancreas are shown in Fig. 4 in this paper and have typical morphology and size[66-68].

Table 3. Dimensional characteristics of the structural components of the mouse pancreas

| Structural components | Calculated dimensions (see Fig. 4), μm | Dimensions in literature, μm | References |
|---|---|---|---|
| Distance between the capillaries | 10-30 | 20-50 | 66 |
| Capillaries | 4-10 | 3-7 | 66-68 |

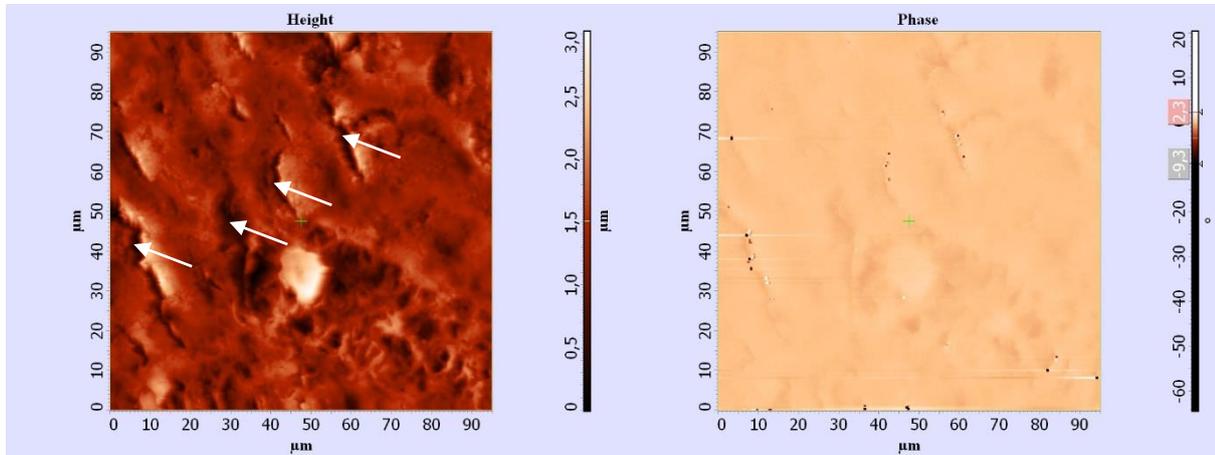

Fig. 4. AFM (left) and MFM (right) images of a sample of the pancreas of the house mouse *Mus musculus*: arrows indicate the capillaries.

Capillaries of the intestine are presented in Fig. 5 in this work and have also typical morphology and size, described in the work[69,70].

Table 4. Dimensional characteristics of the structural components of the mouse intestine

| Structural components | Calculated dimensions (see Fig. 5), μm | Dimensions in literature, μm | References |
|---|---|---|---|
| Distance between the capillaries | 5-30 | 30-40 | 71,72 |
| Capillaries | 3-6 | 3-5 | 69,70 |

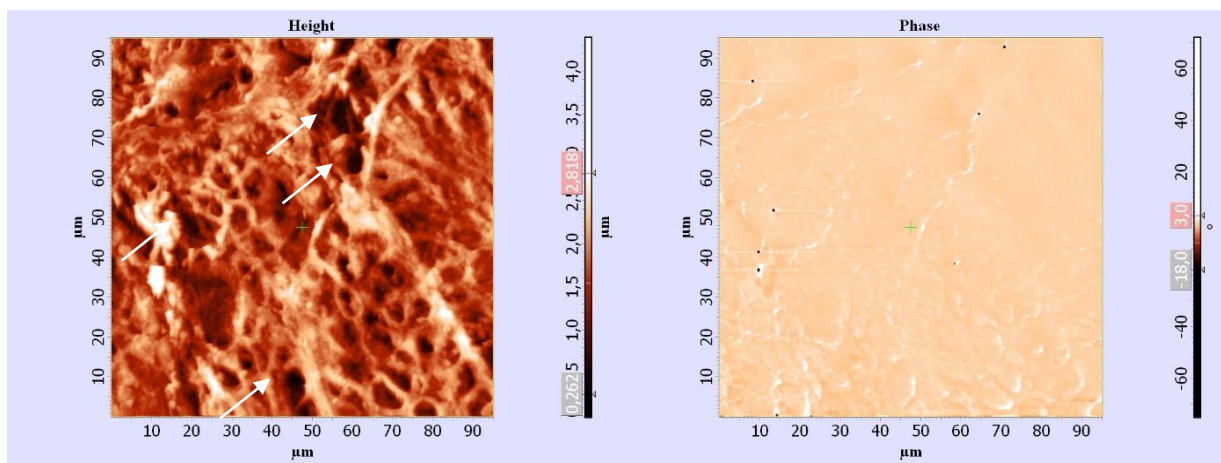

Fig. 5. AFM (left) and MFM (right) images of a sample of the intestine of the house mouse *Mus musculus*: arrows indicate the capillaries.

AFM and MFM examinations of porcine specimens gave similar results, namely, BMNs in the lungs (Fig. 5 and Table 6), kidneys (Fig. 6 and Table 7) and spleen (Fig. 7 and Table 8) of the pig are located in the capillary wall. So, capillaries of lungs, are presented in fig. 6 in this paper and have typical morphology and size, described in the works[73,74].

Table 5. Dimensional characteristics of the structural components of the pig lung

| Structural components | Calculated dimensions (see Fig. 6), μm | Dimensions in literature, μm | References |
|---|---|---|---|
| Distance between the capillaries | 5-25 | 10-20 | 73 |
| Capillary density | 600-900/μm$^2$ | 700-900/μm$^2$ | 75-78 |
| Capillaries | 5-10 | 5-20 | 73,75 |

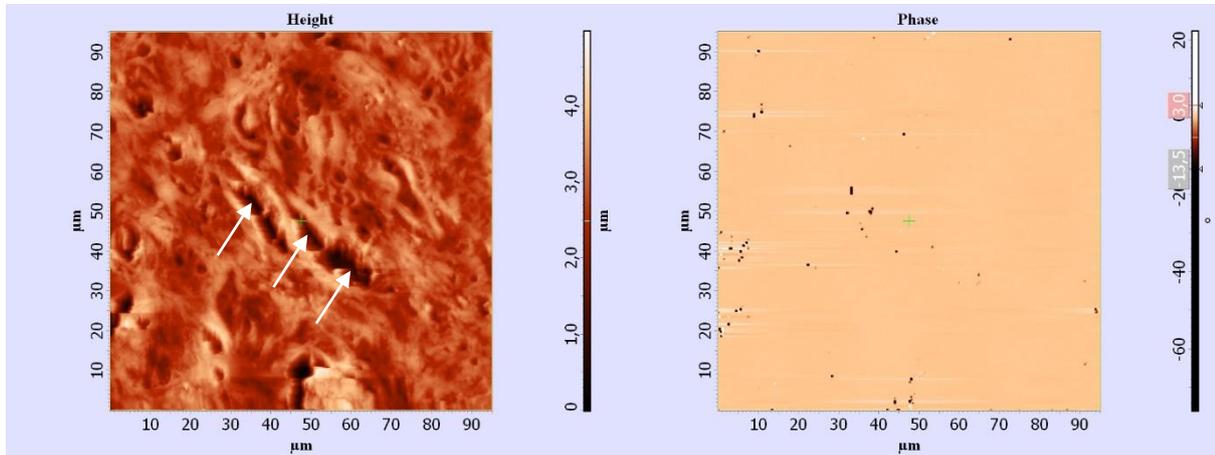

Fig. 6. AFM (left) and MFM (right) images of a sample of the lungs of a domestic pig *Sus domestica*: arrows indicate the capillaries.

The capillaries of the kidneys are presented in Fig. 7 in this paper and have typical morphology and size, described in[79].

Table 6. Dimensional characteristics of the structural components of the pig kidney

| Structural components | Calculated dimensions (see Fig. 7), μm | Dimensions in literature, μm | References |
|---|---|---|---|
| Capillary density | 100-300/μm2 | 25-220/mm$^2$ | 80,81 |
| Capillaries | 7-10 | 3-20 | 82,83 |

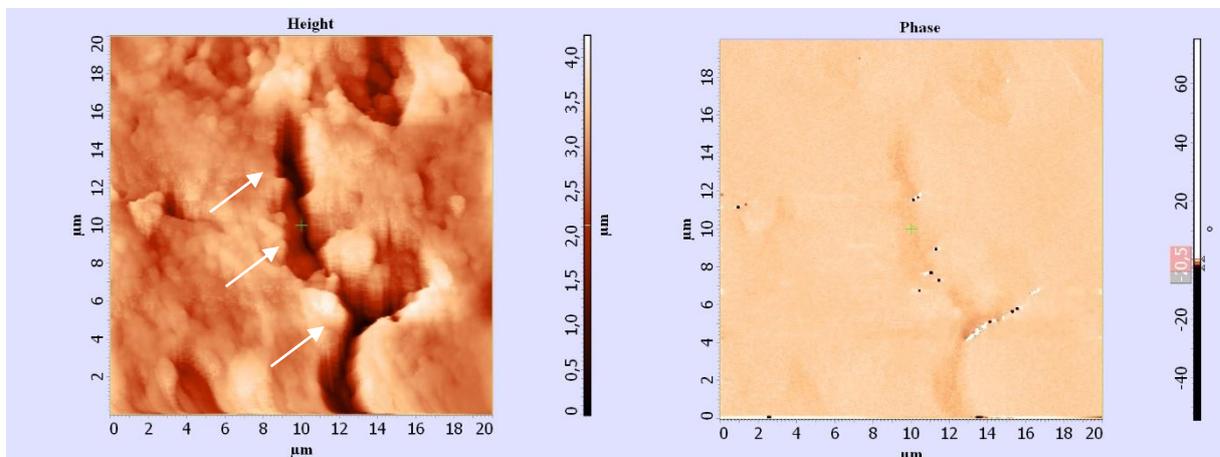

Fig. 7. AFM (left) and MFM (right) images of a sample of thekidney of the domestic pig *Sus domestica*: arrows indicate the capillaries.

The capillaries of a spleen, are presented in fig. 8 in this work, have typical morphology and size, described in[84,85].

Table 7. Dimensional characteristics of the structural components of the pig spleen

| Structural components | Calculated dimensions (see Fig. 8), μm | Dimensions in literature, μm | References |
|---|---|---|---|
| Distance between the capillaries | 5-10 | 10-30 | 86 |
| Capillaries | 4-7 | 4-10 | 87-90 |

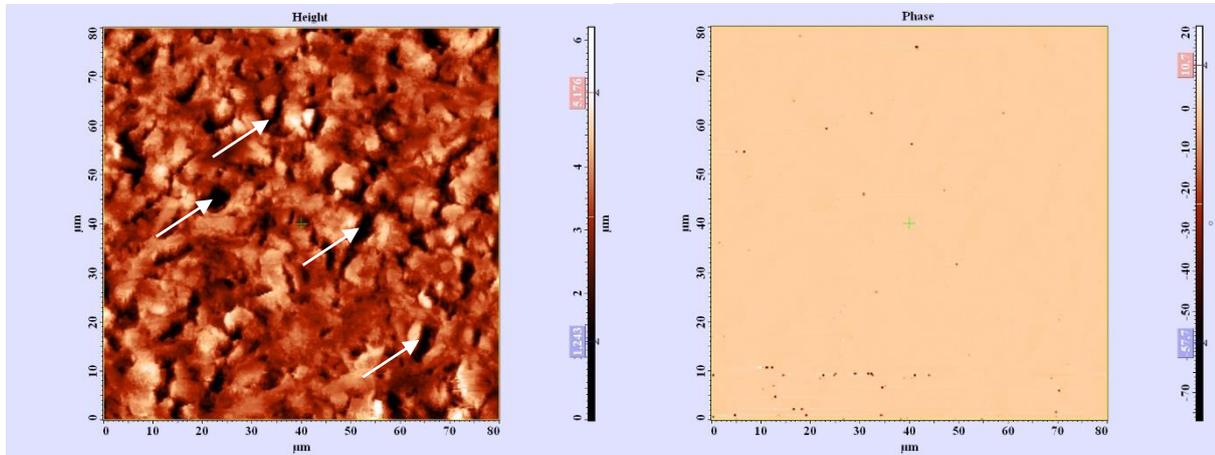

Fig. 8. AFM (left) and MFM (right) images of a sample of the spleen of the domestic pig *Sus domestica*: arrows indicate the capillaries.

As noted above, BMNs are most thoroughly investigated in the organs and tissues of fishes. However, only transmission electron microscopy (TEM) and MFM provide information on the location of BMNs among all experimental methods that are used today to study BMNs[34,35,91-114]. This paper presents the results of AFM and MFM studies of samples of the silver carp brain, and it is shown that BMNs are located in the capillary wall (Fig. 9 and Table 8) in the brain of the silver carp, as in the organs of mice and pigs. The capillaries of brain are presented in fig. 9 in this work and have typical morphology and dimensions, characterized in[115].

Table 8. Dimensional characteristics of the structural components of the silver carp brain

| Structural components | Calculated dimensions (see Fig. 9), μm | Dimensions in literature, μm | References |
|---|---|---|---|
| Distance between the capillaries | 10-30 | 15,3 | 116 |
| Capillaries | 3-9 | 3-8 | 116,117 |

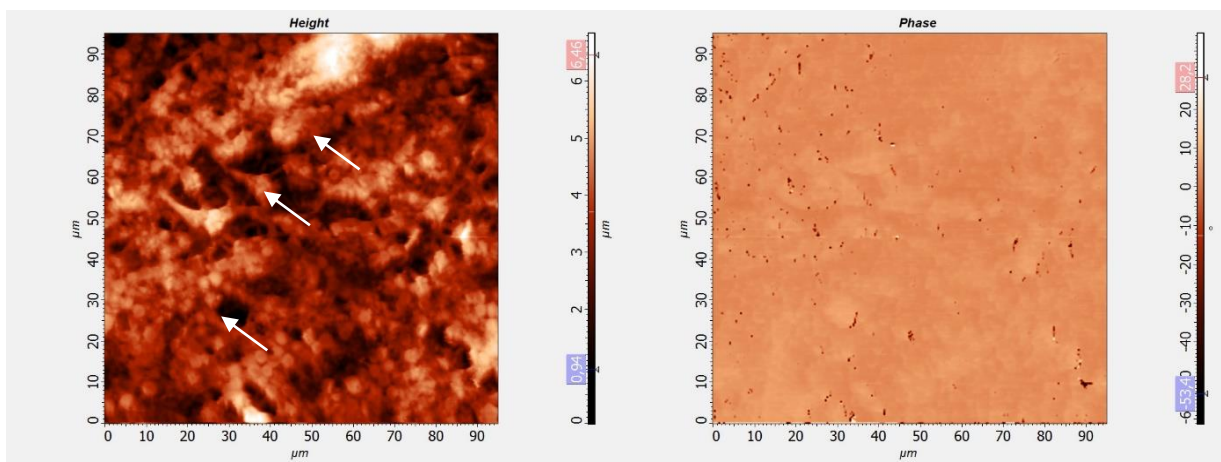

Fig. 9. AFM (left) and MFM (right) images of a sample of the silver carp brain of *Cyprinus carpio*: arrows indicate the capillaries.

The BMNs in the human ethmoid bone are located in the sinuses of the bone[56], through which the capillaries and nerves pass[118-120]. The location of BMNs in the ethmoid bone of fishes is similar[25]. Thus, a common feature of the location of BMNs in the ethmoid bone and in other organs of various animals (Fig. 2-9, Table 1) is their localization in the vicinity of the capillaries.

Due to the fact that for many years, BMNs was investigated mainly in connection with the ideas about magnetotaxis and magnetoreception, BMNs in plants and fungi has been little studied. For example, the paper[11] presents experimental data on the detection of magnetically sensitive nanoparticles in plants, but the authors of this work consider it more likely that they have detected phytoferritin and its aggregates, and not plant BMNs. As for fungi, BMNs is experimentally detected only in the microfungi *Fusarium oxysporum* and *Verticillium spp*.[121].

The study of BMNs in plants was carried out on the example of tobacco, as the most studied model organism among plants, as well as potatoes. The BMNs in the leaf and the roots of tobacco (Fig. 10-11) are located on the membrane of phloem sieve tubes. Phloem is a conducting plant tissue that forms a network of sieve tubes through which organic substances, synthesized by leaves during photosynthesis, are transported to all plant organs[122-126], unlike xylem, which transports water and minerals from the soil[127-130]. Sieve tubes of tobacco leaf are presented in Fig. 10 in this paper and have the typical morphology and dimensions described in[131].

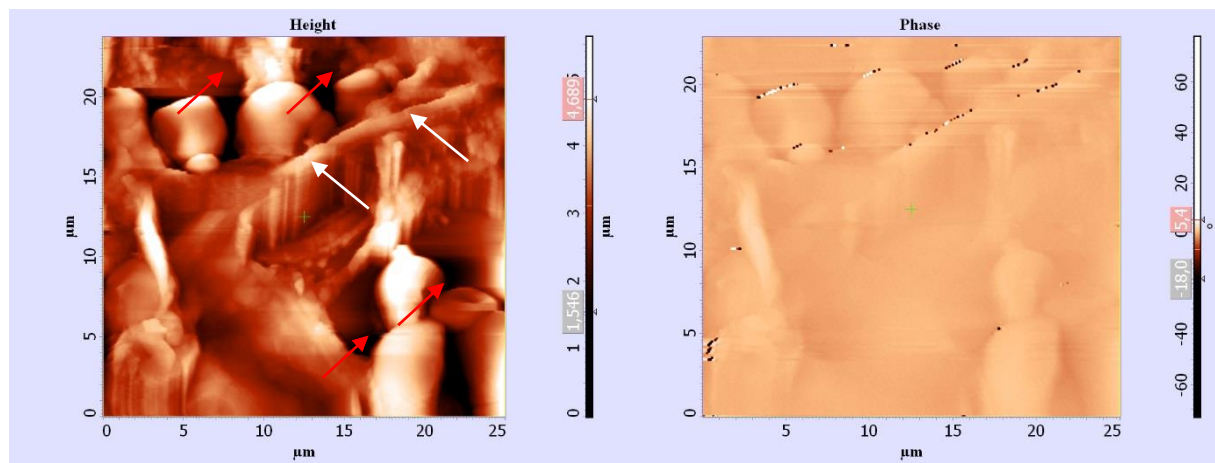

Fig. 10. AFM (left) and MFM (right) images of a sample of a leaf of the cultivated tobacco *Nicotiana tabacum*. Elements of sieve tubes, including pores, are visible in the AFM image. White arrows in the AFM image indicate the membrane of the sieve tubes, red arrows indicate the pores of the sieve tubes.

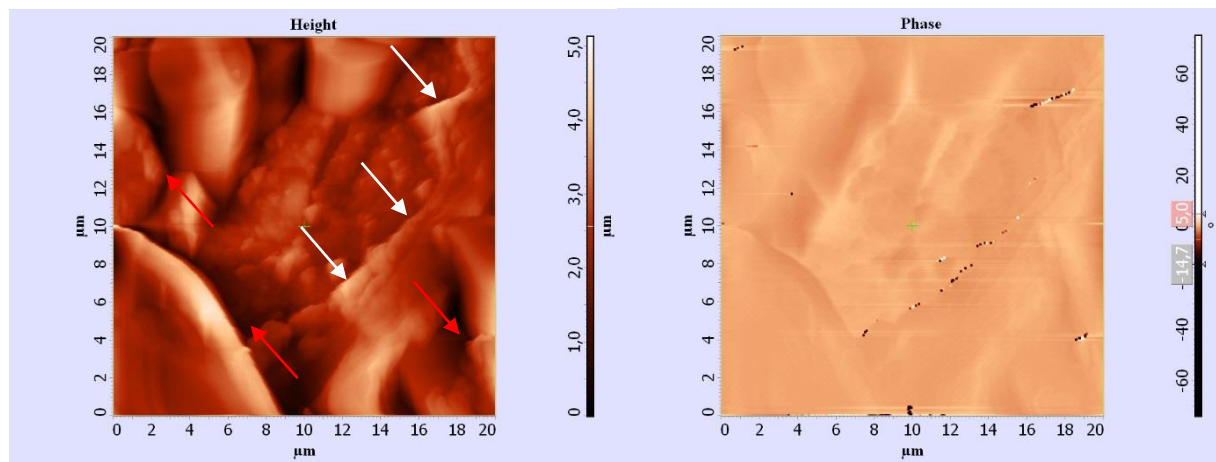

Fig. 11. AFM (left) and MFM (right) images of a sample of a root of the cultivated tobacco *Nicotiana tabacum*. Elements of sieve tubes, including pores, are visible in the AFM image. White arrows in the AFM image indicate the membrane of the sieve tubes, red arrows indicate the pores of the sieve tubes.

Similar data on the location of BMNs were obtained on the samples of potatoes. The BMNs in the potato stalk are located in the walls of the conducting tissue (phloem) (as can be seen from Fig. 12-

13), which is characterized in[132]. BMNs in a potato tuber are located at the boundary between starch grains[133] and conducting tissue (phloem).

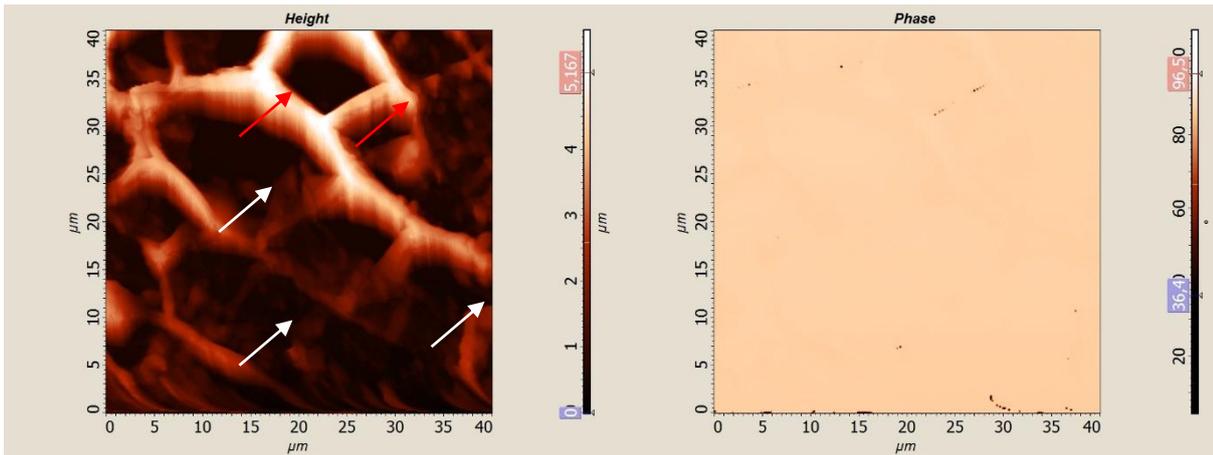

Fig. 12. AFM (left) and MFM (right) images of a sample of the stem of a potato *Solanum tuberosum*: conducting tissue/sieve tubes (white arrows) phloem, cell wall (red arrows).

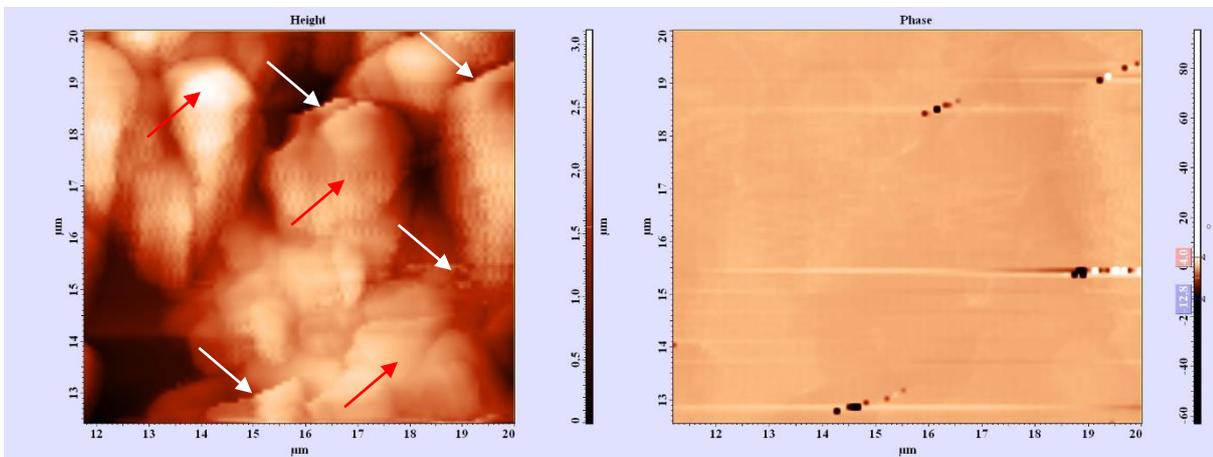

Fig. 13. AFM (left) and MFM (right) images of a sample of the tuber of a potato *Solanum tuberosum*: conducting tissue/sieve tubes (white arrows) phloem, cell wall (red arrows).

The study of BMNs in mushrooms was carried out using the example of higher mushrooms, such as mushrooms *Agaricus bisporus* and *Lentinula edodes*, which are the most common edible mushrooms. As can be seen from Fig. 14-15 BMNs in the mushroom are located in the walls of the vascular hyphae, which are characterized in the works[134-136].

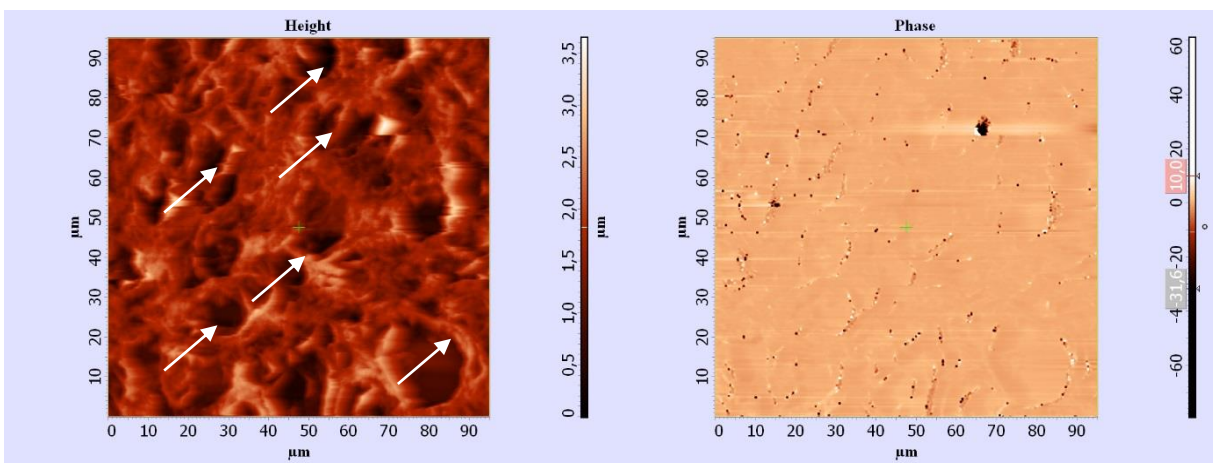

Fig. 14. AFM (left) and MFM (right) images of a sample of the common mushroom *Agaricus bisporus*: arrows indicate the hyphae.

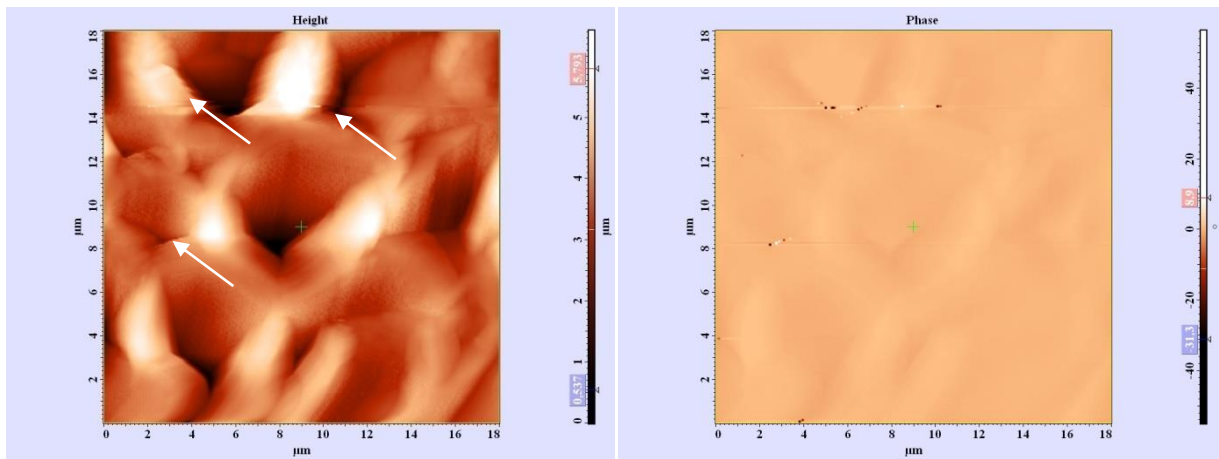

Fig. 15. AFM (left) and MFM (right) images of a sample of the shiitake mushroom *Lentinula edodes*: arrows indicate the hyphae.

Thus, studies of BMNs in samples of organs and tissues of animals, plants and fungi, conducted in this work showed that:
• BMN in the organs of multicellular organisms form chains;
• BMN in multicellular organisms are part of the transport system. Thus, BMNs in animals are located in the walls of capillaries (all examined organs and tissues except the ethmoid bone) or in the vicinity of the capillaries (the ethmoid bone). BMNs in plants are located in the wall of the conducting tissue, namely in the wall of sieve tubes of the phloem. BMNs in fungi are located in the wall of the conducting tissue, namely in the walls of the vascular hyphae[134-136].

**Conclusions**
The chains of BMN are components of the cells that form the walls of the capillaries of animals and the walls of the conducting tissue of plants and fungi. At the same time, the functions of the capillaries of animals and the functions of the conducting tissue of the phloem of plants, as well as the vascular hyphae of fungi, are similar[122,137,138]. In particular, capillaries (exchange vessels) exchange nutrients, gases, liquids, metabolites, signal substances (hormones), immune cells, etc. between the blood and body tissues[137,139-141]. Conducting tissue of the phloem of plants serves to transfer organic substances, hormones, etc. through the body[122-126]. The same functions have vascular hyphae of fungi[138,142-144]. The location of BMNs in various organs and tissues of animals, plants and fungi as part of systems with similar functions cannot be accidental, taking on account that the genetically programmed mechanism of biosynthesis of BMNs arose at the beginning of evolution[42,43,145-147]. Such localization of BMNs argues in favour of the idea that BMNs chains are directly involved in metabolic processes[25,42,43,148-150] and perform vital functions. In addition, bioinformatics analysis showed that BMNs in multicellular organisms, as well as in magnetotactic bacteria[151-154], form inside the cell and are associated with the cell membrane[8,9,15,34,39]. This means that the chains of BMNs in multicellular organisms are part of a new type of organelles with the ferrimagnetic properties – the ferrimagnetic organelles of a specific purpose. Such ferrimagnetic organelles create scattering magnetic fields of the order of several fractions of T and gradients of magnetic field induction of the order of $10^5$-$10^6$ T/m in their vicinity, which can significantly affect the mass transfer processes near the cell membrane of vesicles, granules[41,42,150], organelles, structural elements of the membrane and others.

In connection with the foregoing, it is clear that an extremely important task is to search for the general functions of BMNs in various organs and tissues of multicellular organisms. To solve this problem, it is necessary to identify the types and functions of cells containing BMNs in various organs of multicellular organisms.

**Materials and Methods**
**Atomic force microscopy and magnetic force microscopy.** Determination of the presence of BMNs and the study of their localization in the investigated samples was carried out using the «Solver PRO-M» scanning probe microscope by atomic force microscopy (AFM) and magnetic force microscopy (MFM).

The magnetic probe MFM_LM series with chip size 3.4x1.6x0.3mm, coated by CoCr was used. This probe was used both for AFM and MFM imaging. The non-contact AFM (NC-AFM) mode was applied. The MFM scanning was carried out at the constant distance from the sample surface after AFM scanning. The probe "lift" height was 100 nm. The cantilever was calibrated using the test samples. Calibration of the probe was carried out immediately before the measurements.

The biological material was prepared before AFM and MFM scanning. Fixation of biological material was carried out in a 10% formalin solution. The duration of fixation was 24 hours. After that, the biological material washed in distilled water and conducted through ethanol with increasing concentration (from 50% to 100%). The next stage was the impregnation of the with liquid paraffin at a temperature of 55°C. After the solidification of paraffin at room temperature, a paraffin block was obtained. Slices from a paraffin block 5 μm thick were obtained using a microtome. After receiving the slices, they were placed on slides. The last stage was the release of slices from the mounting medium.